\documentclass[aps,superscriptaddress,showpacs,showkeys]{revtex4}
\usepackage[latin1]{inputenc}
\usepackage[english]{babel} 
\usepackage[]{graphicx}
\usepackage[mathcal]{euscript}
\usepackage{amsmath}
\usepackage{amssymb}
\usepackage{amsthm}
\graphicspath{{./}{fig/}}
\usepackage{epsfig}
\newcommand{\ba}{\begin{eqnarray}}
\newcommand{\ea}{\end{eqnarray}}
\newcommand{\avg}[1]{\langle #1 \rangle}

\newcommand\e{\emph}
\begin{document}
\title{Stochastic analysis of an agent-based model}
\author{A. Veglio}
\email{andrea.veglio@unito.it}
\affiliation{Department of Oncological Sciences, University of Torino and Institute for Cancer Research and Treatment, Candiolo (To), Italy}
\author{M. Marsili}
\email{marsili@ictp.trieste.it} 
\affiliation{The Abdus Salam International Center of Theoretical Physics, Trieste, Italy}
\begin{abstract}
We analyze the dynamics of a forecasting game which exhibits the phenomenon of information cascades. Each agent aims at correctly predicting a binary variable and he/she can either look for independent information or herd on the choice of others. We show that dynamics can be 
analitically described in terms of a Langevin equation and its collective behavior is described by the solution of a Kramers' problem. This provides very accurate results in the region where the vast majority of agents herd, which corresponds to the most interesting one from a game theoretic point of view.
\end{abstract}
\maketitle

\section{Introduction}
Information aggregation plays an essential role in subsistence, evolution and
self-organization of living beings. The selective advantages of sexual over asexual reproduction relies partly on the higher efficiency of the first to aggregate genetic information of fitter individuals~\cite{sex}.
Fishes, birds or mammals join together in shoals to gather
from each other information on food or predators placement~\cite{levin1}. 
In human behaviour, emergence of fashion, fads, social
consensus are expression of (attempts of) information aggregation. In economics, markets have been celebrated as having the virtue of correctly aggregating dispersed information into prices on a theoretical basis, but
the occurrence of crashes and bubbles clearly suggests that in practice markets may fail in this respect \cite{bouchaud}. 

The failure of information aggregation can arise because of herding, as vividly explained be the phenomenon of \emph{rational herding} and  \emph{information cascades}~\cite{bikh2}. In brief, these are situations where many individuals aim at gathering some information about an event, having access to some private information. If all individuals choose according to their private signals, the choice of the majority correctly aggregates all available information.
This means that the choice of the majority conveys much more precise information than the one each agent receives and hence, if agents can observe the choice of others, it becomes profitable to follow the majority's choice. The problem is that if all agents do this, choices no longer reflect private information, and information aggregation miserably fails.

The emergence of large fluctuations in socio-economic phenomena such as social consensus~\cite{marscons}, herding \cite{mars}, bubbles and crashes in financial markets~\cite{crashes,physicaA} can be conveniently formalized in simple Ising like models. This approach reveals clearly the role which (spontaneous) symmetry breaking and long range order have in these phenomena. It also allows, in principle, to derive a precise description of the dynamical process taking place.

In this paper, we study in details a simple model where information cascades arises in a system of spins interacting on a directed graph and which is intimately connected to phase coexistence~\cite{mars}. 
We will show that its dynamics can be analytically described in terms of a Langevin equation and that  collective behavior is related to the solution of a Kramers' problem. This approach yields excellent results in the most peculiar
region of the phase diagram.

\section{The model}

The model describes a population of agents engaged in the task of predicting a binary variable $E=\pm 1$ (e.g. whether something will occur or not). Each agent has two possible strategies, either to look independently for information on the event, or to infer information from their peers. The first option delivers a signal which predict the correct outcome with a probability $p>1/2$ whereas the second option, i.e. if the agent decides to herd, provides that he or she will have access to the predictions of a group of $K>1$ other agents. While informed agents will follow the recommendation of the signal they receive, herders will enter into an information exchange stage, where their predictions will gradually converge to those of the majority of their peer group. The key issue, depending on how many agents will decide to herd, is what the majority of the population will do and, given this, what is the Nash equilibrium strategy, i.e. that which maximizes the probability of predicting the correct outcome. When most agents look for independent information, herding is the best move as it delivers the aggregate information of more than one agent. However, if all agents herd, no one will look for information suggesting the population will not be able to predict the correct outcome.

In order to model such a situation, Ref. \cite{mars} considers a system of $N$ agents, each characterized by a spin variable $s_i$, taking the value $s_i=+1$ in the case of a correct prediction and $s_i=-1$ otherwise. The information exchange process can be described by the following asynchronous update process: At each discrete time $n=0,1,\ldots$ an agent $i$ is picked at random and his/her spin is updated as
\begin{equation}
\label{upd}
s_i(n+1)={\rm sign}\left[h_i+\sum_{j=1}^N a_{i,j}s_j(n)\right]
\end{equation}
whereas $s_j(n+1)=s_j(n)$ for all $j\neq i$. We denote with $i\in I$ informed agents and with $i\in H$ herders. The difference in the two strategies is encoded in the values of $h_i$ and $a_{i,j}$. 
The former describes the signals received by informed agents
\begin{equation}
\label{hi}
h_i=\left\{\begin{array}{cl}
+1 & \hbox{with~probability}~p,~\hbox{for}~i\in I \\
-1 & \hbox{with~probability}~1-p,~\hbox{for}~i\in I  \\
0 & i\in H\end{array}\right.
\end{equation}
wheres $a_{i,j}$ describes the network through which herders collect their information. Therefore, if $G_i$ is the group of agents that agent $i\in H$ copies,
\begin{equation}
\label{aij}
a_{i,j}=\left\{\begin{array}{cl}
1 &i\in H,~j\in G_i \\
0 & \hbox{else}\end{array}\right.
\end{equation}
Each herder $i\in H$ has $|G_i|=K=\sum_j a_{i,j}$ peers in his/her group which are chosen at random among the $N-1$ other agents \footnote{This implies that herder $i$ does not know whether their peers will be informed or not, before selecting them in their group $G_i$. Also note that with this rule, $j\in G_i$, i.e. $i$ observing $j$, does not necessarily  imply $i\in G_j$, i.e. that $j$ observes $i$: this produces a non symmetric type of interaction.}. We shall assume also that $K$ is odd, so that Eq. (\ref{upd}) always provides a well defined recommendation. Summarizing, the first term in the braces of Eq. (\ref{upd}) is non-zero only for $i\in I$ whereas the second is non-zero only for $i\in H$. All spins are initially assigned a random value $s_i(0)=\pm 1$ with equal probability. 

\begin{figure}
\centering
\begin{picture}(0,0)
\includegraphics[scale=0.7]{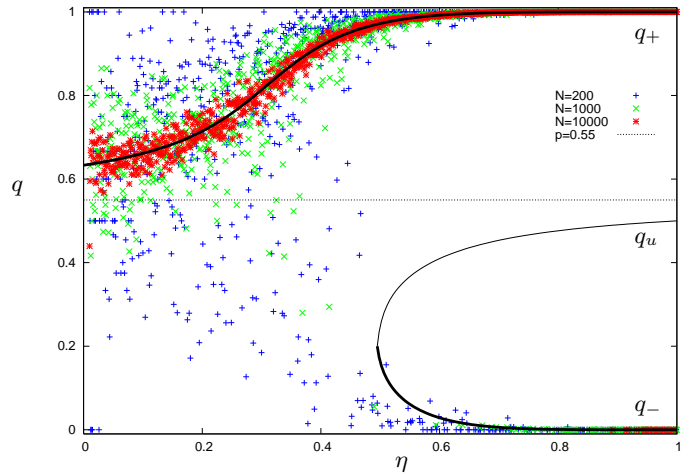}
\end{picture}
\setlength{\unitlength}{3947sp}
\begin{picture}(4245,3105)(-44,-2266)
\put(-264,-576){\makebox(0,0)[lb]\mbox{$q$}}
\put(2146,-2311){\makebox(0,0)[lb]\mbox{$\eta$}}
\put(3650,-900){\makebox(0,0)[lb]\mbox{$q_u$}}
\put(3650,380){\makebox(0,0)[lb]\mbox{$q_+$}}
\put(3650,-1960){\makebox(0,0)[lb]\mbox{$q_-$}}
\end{picture}\
\caption{Probabilities $q$ of a herding agent making the right forecast
  at different fractions $\eta$ of herding agents for $p=0.55$, 
   $K=11$ and $N=200\ (+), 10^3\ (\times), 10^4\
  (\star)$. Full lines represent solutions of 
  $q=\Omega_K(\pi)$ as in~(\ref{igr}): unstable
  solution $q_u$ is plotted through a thiner line.}
\label{disp}
\end{figure}

Fig. \ref{disp} shows the result of numerical simulations for different systems sizes $N$ and different fractions
\begin{equation}
\label{eta}
\eta=\frac{|H|}{N}
\end{equation}
of herding agents. The dynamics converges to a fixed point and we denote simply by $s_i$ the value the $i^{\rm th}$ spin attains.
Fig. \ref{disp} reports the probability of a right forecast of a randomly chosen agent $i\in H$, which is given by
\begin{eqnarray}\label{lnl}
q\equiv\frac{1}{\eta N}\sum_{i\in H}\delta_{s_i,1}
\end{eqnarray}
Each point represents a single realization.

While in the left part of the figure $q$ values are spread around a single value, on the right they are polarized onto two
values, $q\sim 0$ (almost certainly wrong forecast) and $q\sim 1$ 
(almost certainly right forecast). The average $\avg{q}$ of $q$ over different realizations (see Fig. \ref{second}) exhibits a non-monotonic behavior with $\eta$ and a strong dependence on the system size $N$ for $\eta$ close to $1$.

One particularly relevant point $\eta^*$ is that for which $\avg{q}=p$, i.e. where the probability of success is independent of the choice of the strategy. For $\eta>\eta^*$ informed agents perform better than herders. It is reasonable that some herders would switch strategy, thus decreasing $\eta$. Likewise, for $\eta<\eta^*$ informed agents have incentives to switch to herding, thus suggesting an increase in $\eta$. Therefore we expect the population should converge to a fraction $\eta^*$ of herders, which can be thought as a Nash equilibrium.

\section{Mean field in the stationary state}

Fig. \ref{disp} suggests that the system exhibits a
phase transition around $\eta\sim1/2$. 
In order to investigate this behavior, Ref. \cite{mars} suggested a self-consistent approach.
First of all, let us notice that the probability $\pi$ for a 
generic agent (whether informed or
herding) making the right forecast is
\begin{eqnarray}\label{tot}
\pi&\equiv& P(f_i=E\ |\ i\in I\cup H)=(1-\eta)p+\eta q
\end{eqnarray}
Besides, the probability $q$ can be viewed as the sum of probabilities
that, for $i\in H$, every combination of majority in $G_i$ gives rise
to the right forecast, i.e.
 \begin{eqnarray}
q=\sum_{g=\frac{K+1}{2}}^{K}\
\binom{K}{g}\pi^g(1-\pi)^{K-g}\equiv\Omega_K(\pi)\ ,
\label{igr}
\end{eqnarray}
Since~(\ref{tot}) holds, (\ref{igr}) represents an implicit equation in
$q$, whose solutions are plotted with full lines in Fig.~\ref{disp} for
comparison with numerical results. Eq. (\ref{igr}) admits one solution for $\eta<\eta_c$, while three for $\eta>\eta_c$. Individual outcomes of the game cluster around the upper solution $q_+$ of Eq. (\ref{igr}) for $\eta<\eta_c$, whereas some realizations also converges to the lower solution $q_-$ for $\eta>\eta_c$. No realization converges to the intermediate solution $q_u$ for $\eta>\eta_c$. This suggests that the upper and lower branches of the full lines 
correspond to two stable solutions ($q_+$ and $q_-$), whereas the middle one $q_u$ corresponds to an unstable 
solution which separates the basin of
attraction of $q_+$ and $q_-$. This hypothesis is confirmed by the detailed analysis of the dynamics discussed in next section. 

\section{Stochastic dynamics}

\begin{figure}
\centering
\begin{picture}(0,0)
\includegraphics[scale=0.7]{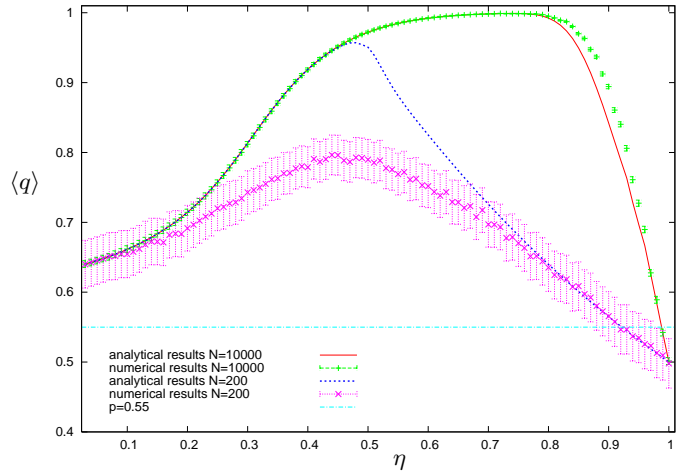}
\end{picture}
\setlength{\unitlength}{3947sp}
\begin{picture}(4245,3105)(-44,-2266)
\put(-264,-576){\makebox(0,0)[lb]\mbox{$\langle q\rangle$}}
\put(2146,-2311){\makebox(0,0)[lb]\mbox{$\eta$}}
\end{picture}
\caption{Average probability $\langle q\rangle$ 
of a herding agent making the right forecast
  at different fractions $\eta$ of herding agents for $p=0.55$, 
$K=11$ and $N=200\ (\times)$ or $10^4\ (+)$.}
\label{second}
\end{figure}

In order to understand the behavior of the average outcome $\avg{q}$ shown in Fig.~\ref{second}, it is necessary to compute the probability that the system will converge to one or the other fixed point $q_\pm$ for $\eta>\eta_c$. Indeed we can write 
\begin{equation}\label{qmean}
 \langle q\rangle=p_-q_-+(1-p_-)q_+
\end{equation}
where $p_-$ is the probability that a realization of the system converges to the
fixed point $q_-$. Eq.~(\ref{qmean}) represents an approximation
for large $N$ since, as shown in Fig.~\ref{disp}, fixed points values deviate from
$q_\pm$ solutions for finite $N$. The dispersion of fixed points is however very small for $\eta$ close to one, which is the relevant region where the Nash equilibrium is located.\\

While  $q_-$ and  $q_+$ are given
by the solution of Eq. (\ref{igr}), in order to derive an expression for $p_-$ 
we will model the dynamics of $q$ as the information exchange process takes place. As we shall see, $q$ can be considered 
as the coordinate of a particle subject to a one dimensional double well potential, with a minimum in $q_+$ and one in
$q_-$, and submitted to a stochastic dynamics. These types of problems,
which are known under the name of
Kramers' problems (e.g.,~\cite{gard}), 
are well known in the theory of stochastic
processes. In order to apply these results, we shall first derive a
continuous time dynamics for $q(t)$ (Langevin equation) 
and then use the backward
Fokker-Planck equation to evaluate the probability $\tilde{p}_-(q_0)$
that a single realization of the 
system, starting from an initial
``position'' $q_0$
and evolving according to a stochastic
dynamics, falls in the stable solution $q_-$.\\

We stress that we denote by $\tilde{p}_-(q_0)$ the probability that a {\it single}
realization of the system, with initial condition $q_0$ falls in $q_-$ to distinguish it from
\begin{equation}
\label{pmeno}
p_-=\int_0^1dq_0\ \tilde{p}_-(q_0)\ \mathcal{R}(q_0),
\end{equation}
i.e. from the probability that \emph{a generic} realization falls in $q_-$, which is obtained taking the average over the distribution $\mathcal{R}(q_0)$ of $q_0$. For each realization of the system, the initial value $q_0$ depends on the initial draw of the spins $s_i(0)$ and hence, by the central limit theorem, the probability distribution of $q_0$ is well approximated by a Gaussian with mean $1/2$ and variance $1/(\eta N)$, i.e. 
\begin{equation}
\label{Rq0}
\mathcal{R}(q_0)\equiv\sqrt{\frac{2\eta N}
{\pi}}e^{-2\eta N(q_0-\frac{1}{2})^2}
\end{equation}

In order to derive a continuum time description of the problem, we introduce a continuous time $\tau$ s.t. $\tau\equiv\frac{1}{\eta N}n$ and consider the variation $dq=q(\tau+d\tau)-q(\tau)$ over a time interval $d\tau=\frac{\Delta n}{\eta N}$. In a single time step, we have $q(n+1)=q(n)+\frac{1}{2\eta
  N}\left[s_{i(n)}(n+1)-s_{i(n)}(n)\right]$, where $i(n)$ stands for the selected
herding agent. Summing this over $n\in[\eta N\tau,\eta N(\tau+d\tau))$ and using Eq. (\ref{upd}) yields
\begin{equation}\label{clt}
q(\tau+d\tau) = q(\tau) + \frac{1}{2\eta N}
\sum_{n=\eta N\tau}^{\eta N\tau+\Delta n-1}
\bigg(\mbox{sign}\sum_{j\in G_{i(n)}}s_j(n)-s_{i(n)}(n)\bigg).
\end{equation}
For $N\gg 1$ we can have $d\tau\ll 1$, as appropriate for deriving continuum time equations, but with $\Delta n\gg 1$, which allows us to apply the CLT to estimate the right hand side. In order to do this, we regard $s_i(n)$ as random variables with distribution 
\[
P(s_i(n)=+1|i\in H)=q(\tau), ~~~~n\in [\eta N\tau,\eta N\tau+\Delta n)
\]
if $i\in H$, whereas $P(s_i(n)=+1|i\in I)=p$. This allows us to estimate the mean and the variance of the right hand side of Eq. (\ref{clt}). Therefore, for $N\gg 1$ 
the dynamics is well aproximated by the Langevin equation
\begin{equation}
\frac{dq(\tau)}{d\tau}=b\big[q(\tau)\big]+
\sigma\sqrt{a\big[q(\tau)\big]}\ \xi(\tau)
\end{equation}
where $\xi(\tau)$ is a white noise term with $\langle\xi(\tau)\rangle=0$ and    
$\langle\xi(\tau)\xi(\tau')\rangle=\delta(\tau-\tau')$ and whose 
drift and diffusion terms are given by
\begin{eqnarray}
b[q(\tau)]&=&\Omega_K(\pi)-q(\tau)\label{pos1}\\
a[q(\tau)]&=&\Omega_K(\pi)\big[1-\Omega_K(\pi)\big]+q(\tau)\big[1-q(\tau)\big]
\label{pos2}\\
\sigma^2&=&1/(\eta N)\label{pos4}
 \end{eqnarray}

In passing, it is worth to remark that $q_u$ is indeed an unstable fixed point of the deterministic dynamics, while $q_\pm$  are stable. In fact, denoting the first derivative of $b[q]$ in $q$
 by $b'[q]$  and the generic dynamics' fixed point
by $q^*$, it is easy to show that $b'[q^*]>0$ if $q^*=q_u$ and $b'[q^*]<0$ if
$q^*=q_\pm$. Therefore, considering the linear expansion of $q$ around
$q^*$, $q=q^*+\delta q$, and the first order relationship $\delta\dot{q}\simeq
\frac{\partial\dot{q}}{\partial{q}}\big|_{q=q^*}\delta q\simeq 
b'[q^*]\delta q$, we see that $q_u$ is unstable while $q_\pm$
are stable. 

Let us now turn to the computation of   
$\tilde{p}_-(q_0)\equiv {P(q_-,\tau=\infty\ |\ q_0,\tau_0=0)}$, i.e. 
the probability that the system reaches $q_-$ starting from $q_0$ 
in a time interval $\Delta\tau=\infty$. As a function
of $q_0$ and $\tau_0$, $\tilde{p}_-(q_0)$ satisfies a backward Fokker-Planck
equation
\begin{eqnarray}\label{yoghi}
b[q_0]\ \partial_q \tilde{p}_-(q)\big|_{q=q_0}+\frac{\sigma^2}{2}\ a[q_0]
\ \partial_{q}^{2}\tilde{p}_-(q)\big|_{q=q_0}=0
\end{eqnarray}
whose drift and diffusion coefficients $b[q_0]$ and $\sigma\sqrt{a[q_0]}$ 
are given by~(\ref{pos1}),
(\ref{pos2}), (\ref{pos4}) evaluated at $\tau=0$
and whose boundary conditions are chosen to be absorbing, i.e.
$\tilde{p}_-(q_0)\big|_{q_0=q_-}=1$ and 
\mbox{$\tilde{p}_-(q_0)\big|_{q_0=q_+}=0$}.\\ 
Defining
\begin{eqnarray}
 \varphi(q)\equiv \mbox{exp} \left[\ \frac{2}{\sigma^2}
  \int_{{q_-}}^{q} dq'\  
\frac{ b[q']}{a[q']}\ \right]\ ,\label{phix}
\end{eqnarray}
the solution to Eq.~(\ref{yoghi}) is given by 
$\tilde{p}_-(q_0)=\int_{q_0}^{q\sigma_+} dq\ 
\varphi(q)/\int_{q_-}^{q_+} dq\ \varphi(q)$.\\
Therefore, from Eq. (\ref{pmeno}), the probability that a generic realization of the process converges to the fixed point $q_-$ is given by
\begin{eqnarray}\label{pminus}
p_-=\sqrt{\frac{2\eta N}{\pi}}\int_0^1 dq_0\ \frac{\int_{q_0}^{q_+} dq\ 
\varphi(q)}{\int_{{q_-}}^{q_+} dq\ \varphi(q)}\ 
e^{-2\eta N(q_0-\frac{1}{2})^2}\ .
\end{eqnarray}
We numerically integrated~(\ref{pminus}) with Gauss-Legendre
method. Results are shown in Fig.~\ref{second}.\\
The analytical results are in excellent agreement with numerical simulations for finite $N$ in the region around Nash equilibrium point ($\eta$ close to one).\\
The lower accuracy of the analytic approach in the intermediate $\eta\approx 1/2$ region is probably due to the larger dispersion of fixed points for finite $N$, which is neglected here. Note also that $q_\pm$ are not absorbing points of the dynamics as long as $q_\pm\neq 0$ or $1$, because while the deterministic term vanishes the stochastic term of Eq. (\ref{pos2}) is non-zero. Again, for $\eta$ close to one, this effect is negligible as indeed $q_-\cong 0$ and $q_+\cong 1$. 

\section{Conclusions}

In conclusion, we derived an accurate approximation for the stochastic dynamics of a simple model that emulates a phenomenon of 
information aggregation. 
A static analysis of the model suggests the presence of a phase transition, characterized by phase coexistence of three
solutions, two stable and one unstable. We have shown that the dynamics can be described analytically as a Langevin
equation and the problem of computing the probability to end up in one of
the two stable equilibrium points can be cast in the form of a Kramers' problem.
This approach describes excellently the system in the region
where the vast majority of the agents herd, which is the most interesting one from the point of view of game theory, i.e. around the Nash equilibrium point. \\

Open problems are a more fitting description of the system in region 
\mbox{$0\ll\eta\ll\eta^*$}, the analysis of system's behaviour in case of (partially) symmetric interaction or the effects of contrarian agents (anti-ferromagnetic interaction).

\end{document}